\documentclass{article}

\usepackage{arxiv}

\usepackage[utf8]{inputenc} 
\usepackage[T1]{fontenc}    
\usepackage{hyperref}       
\usepackage{url}            
\usepackage{booktabs}       
\usepackage{amsfonts}       
\usepackage{nicefrac}       
\usepackage{microtype}      
\usepackage{lipsum}		
\usepackage{graphicx}
\usepackage{natbib}
\usepackage{doi}

\title{Comparing Conventional and Conversational Search Interaction using Implicit Evaluation Methods \thanks{Kaushik, A. and J. F. Jones, G. (2023). Comparing Conventional and Conversational Search Interaction Using Implicit Evaluation Methods. In Proceedings of the 18th International Joint Conference on Computer Vision, Imaging and Computer Graphics Theory and Applications - HUCAPP, ISBN 978-989-758-634-7; ISSN 2184-4321, SciTePress, pages 292-304. DOI: 10.5220/0011798500003417}}

\author{ \href{https://orcid.org/0000-0002-3329-1807}{\includegraphics[scale=0.06]{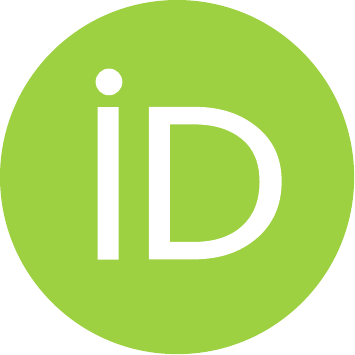}\hspace{1mm}Abhishek Kaushik}\thanks{Now at Dundalk Institute of Technology, Dundalk, Ireland.} \\
	ADAPT Centre, School of Computing \\ 
  Dublin City University \\
  Dublin 9, Ireland \\
	\texttt{abhishek.kaushik2@mail.dcu.ie} \\
	\And
	\href{https://orcid.org/0000-0003-2923-8365}{\includegraphics[scale=0.06]{orcid.pdf}\hspace{1mm}Gareth J. F. Jones} \\
 	ADAPT Centre, School of Computing \\ 
  Dublin City University \\
  Dublin 9, Ireland \\
	\texttt{Gareth.Jones@dcu.ie} \\
}

\renewcommand{\headeright}{}
\renewcommand{\undertitle}{}
\renewcommand{\shorttitle}{}

\begin{document}
\maketitle

\begin{abstract}
	Conversational search  applications offer the prospect of improved user experience in information seeking via agent support. However, it is not clear how searchers will respond to this mode of engagement, in comparison to a conventional user-driven search interface, such as those found in a standard web search engine. We describe a laboratory-based study directly comparing user behaviour for a conventional search interface (CSI) with that of an agent-mediated multiview conversational search interface (MCSI) which extends the CSI.
User reaction and search outcomes of the two interfaces are compared using implicit evaluation using five analysis methods:
claiming to have a better search experience in contrast to a corresponding standard search interface.
\end{abstract}


\section{Introduction}
\label{Introduction}
The growth in networked information resources has seen search or information retrieval become a ubiquitous application, used many times each day by millions of people in both their work and personal use of the internet. For most users, their experience of search tools is dominated by their use of web search engines, such as those provided by {\em Google\/} and {\em Bing\/}, on various different computing platforms. 
Users lack of knowledge on the topic of their information need often means that they must perform multiple search iterations. This enables to learn about their area of investigation and eventually 
to create a query which sufficiently describes their information need which is 
able to retrieve relevant content. The search process is thus often cognitively demanding on the user and inefficient in terms of the amount of work that they are required to do.


Bringing together the needs of users to search unstructured information technologies and advances in artificial intelligence, recent years have 
seen rapid growth in research interest in the topic of {\em conversational search (CS)\/} systems \cite{radlinski2017theoretical}. 
CS systems assume the presence of an agent of some form which enables a 
dialogue-based interaction between the searcher and the search engine to support the user in satisfying their information needs \cite{radlinski2017theoretical}. 
Studies of CS to date have 
generally adopted a human ``wizard'' in the role of the search agent \cite{Trippas:2017:PIC:3020165.3022144,avula2018searchbots}. 
These studies have been conducted in CS systems with the implicit assumption that an agent can interpret the searcher's actions with human like intelligence. In this study, we take a alternative position using an automatic rule-based agent to support the searcher in the CS interface and compare this with the effectiveness of a similar CSI to perform the same search tasks. In this study, we introduce a desktop based prototype MCSI to a search engine API 
Our interface combines a CS assistant with an extended standard graphical search interface. 
The goals of our study include both better understanding of how users respond to CS interfaces and automated agents, and how these compare with the user experience of a CSI for the same task.

The ubiquity of CSIs means that users have well established mental models of the search process from 
their use of 
these tools. With respect to this, it is important to consider that it has been found in multiple studies that subjects find it difficult to adapt to new technologies, especially when dealing with interfaces \cite{krogsaeter1994user}. Thus, when presented with a new type of interface for an equivalent search task, it is interesting to consider how users will adapt and respond to it.

Previous studies of CS interfaces have focused on chatbot type interfaces which limit the information space  of the search \cite{avula2018searchbots,avula2020wizard}, and are very different from conventional graphical search interfaces. Search via engagement with a chat type agent can result in the development of quite different information-seeking mental models  to those developed in the use of standard search systems, meaning that it is not possible to directly consider the potential of CS in more conventional search settings based on these studies.
We are interested in this study to consider how user mental models of the search process from CSIs will response in a CS conversational setting to enhance the user search experience.

For our study of conversational engagement with a search engine and contrasting it with more conventional user-driven interaction, we adopt a range of implicit evaluation methods. Specifically we use cognitive workload-related factors (NASA Load Task) \cite{hart1988development}, psychometric evaluation for software \cite{lewis1995ibm}, knowledge expansion \cite{wilson2013comparison} and search satisfaction \cite{Abhi}. Our findings show that users exhibit significant differences in the above dimensions of evaluation when using our MCSI and a corresponding CSI.

The paper is structured as follows: Section \ref{recent} overviews existing work in conversational engagement and its evaluation, Section \ref{method} describes the methodology for our investigation, Section \ref{procedure} provides details of our experimental procedure and our results and includes analysis, findings and hypothesis testing and Section \ref{conclude} concludes.

\section{Related work}
\label{recent}
In this section, we provide an overview of existing related work in conversational interfaces, conversational search and relevant topics in evaluation. 

\subsection{Conversational Interfaces}
Conversational interaction (CI) with information systems is a longstanding topic of interest in computing. However, activity has increased greatly in recent years. The key motivation for examining CI is the 
development of interactive systems which enable users to achieve their objectives using a more natural mode of engagement than cognitively demanding traditional user-driven interfaces. 
Such user-driven interfaces require users to develop 
mental models to use them  reliably. Recent research on CI has focused on multiple topics including mode of interaction, the intelligence of conversational  agents, the structure of conversation, and dialogue strategy \cite{mctear2016conversational,Kader2015,roller2020recipes}. Progress in CI can be classified in four facet areas: smart interfaces, modeling conversational phenomena, machine learning approaches, and toolkits and languages \cite{singh2019,braun2019,araujo2020conversational}. 

Current chatbot interfaces have evolved, in common with many areas, from rule-based systems to the use of data driven approaches using machine learning and deep learning methods \cite{nagarhalli2020review}. Toolkits have been developed to support 
the construction and testing of chatbot agents for particular applications. The majority of research on conversational agents has focused on question answering and chit chat (unfocused dialogue) systems. Only very limited work has been done on information-seeking bots, dating mainly from the early 1990s \cite{stein1993conversational}. 
One recent example of a multimodal conversational search is presented in our earlier work \cite{CHII2020Abhi}. This enables 
a user to explore long documents using a multi-view interface. Our current study is focused on evaluation of this interface in comparison to a CSI.

\subsection{Conversational Search}
While users of search tools have become accustomed to standard ``single shot'' interfaces, of the form seen in current web search engines, interest in the potential of alternative conversational search-based tools has increased greatly in recent years \cite{radlinski2017theoretical}. Traditional search interfaces have significant challenges for users, in requiring them to express their information needs in fully formed queries, although users have generally learned to use them to good effect. The idea of agent-support conversational-based interaction supporting them in the search process is thus very attractive. 
Multiple studies have been conducted to investigate the potential of conversational search in different dimensions. These studies however have generally involved use of a human in the role of an agent wizard \cite{avula2020wizard,avula2018searchbots,avula2019embedding}. 
These have the limitation of assuming both human intelligence and error free speech recognition, which will generally not be the case in a real system \cite{Trippas:2017:PIC:3020165.3022144,Trippas2018}. 

Some studies on conversational search have been based completely on a data-driven approach using machine learning methods to extract a query from multiple utterances. The drawback of this approach is that the dialogues are not analyzed based on incremental learning over multiple conversations \cite{nogueira2017task,bowden2017combining}. Other types of studies have developed agents by using an intermediate approach in which a combination of rules is used to form a dialogue strategy from users search behaviour \cite{de1994information} \cite{Abhi}, which guide the user in conversations with the support of a pretrained machine learning model to extract the intent and entities from utterance. 
We follow this last approach in our multiview prototype to understand the user search experience in a conversational setting.

\subsection{Evaluation}
Currently, there is no standard mechanism for evaluation of conversational search interfaces. 
In this 
study, 
we adopt implicit measures in five dimensions \cite{Abhievaluation} user search experience \cite{Abhi}, knowledge gain \cite{wilson2013comparison}, cognitive and physical load \cite{hart1988development}, user interactive experience \cite{UEQ} and usability of the interface software \cite{lewis1995ibm}.

\section{Methodology}
\label{method}
In this section, we describe the details of our user study which aims to enable us to observe and better understand and contrast the behaviour of searchers using a CSI and our prototype MCSI. This section is divided into two subsections: interface design and experimental setup.

\subsection{Prototype Conversational Search System}

\begin{figure*}[!ht]
  \includegraphics[scale=0.46]{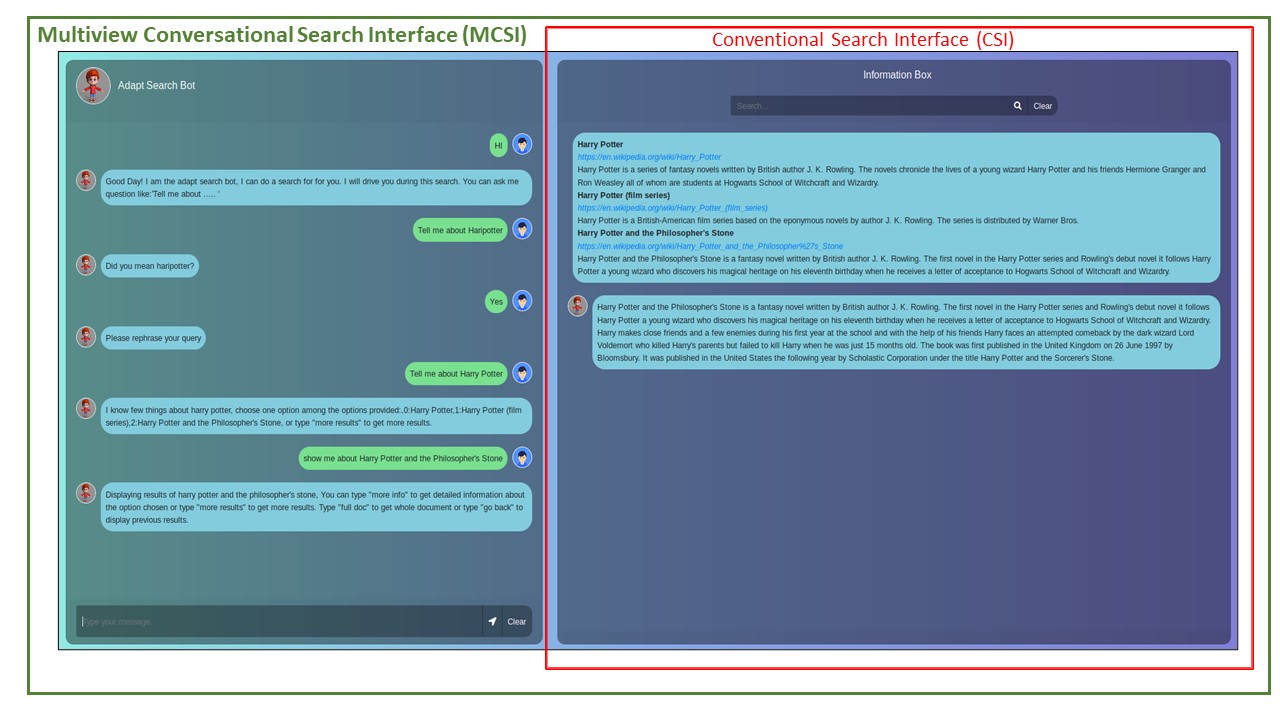}
 \centering
  \caption{Conversational Agent incorporating:
  chat display, chat box, information box, query box with action buttons for Enter and Clear, and retrieved snippets and documents. Green outline indicates the MCSI setting and red block indicates the CSI setting.}
  \label{GUI}
\end{figure*}

In order to investigate user response to search using a MCSI and to contrast this with a comparable CSI 
with the same search back-end, we developed a fully functioning prototype system, shown in Figure \ref{GUI}.
The interface is divided into two distinct sections. The righthand side which corresponds to a standard CSI, 
and the lefthand side which is a text-based chat agent and 
interacts with both the search engine and the user. Essentially the agent works alongside the user as an assistant, rather than being positioned between the user and the search engine \cite{Maes:1994:ARW:176789.176792}.  

The Web interface components are implemented
using the web python framework flask and with HTML, CSS, and JS toolkits. The agent is controlled by a logical system and is implemented using Artificial Intelligence Markup Language (AIML) scripts. These scripts are used to identify the intent of the user, to access a
spell checking API \footnote{https://pypi.org/project/pyspellchecker/}, and are responsible for search and giving responses to the users. 
Since the focus of this study is on the functionality of the search interface, the search is carried out by making calls to the Wikipedia API.
The interface includes multiple components, as discussed in detail in our previous work \cite{CHII2020Abhi}

\subsubsection{Dialogue Strategy and Taxonomy}
After exploring
user search behaviour \cite{Abhi} and dialogue systems, 
we developed a dialogue strategy and 
taxonomy 
to support 
CS. The dialogue process is divided into three phases and four states as discussed in detail in our previous work \cite{CHII2020Abhi} 
The 
three phases include: 
\begin{itemize}
    \item Identification of the information need of the user,
    \item Presentation of the results in the chat system,
    \item Continuation of the dialogue until the user is satisfied or aborts the search.
\end{itemize}

The agent can seek confirmation
from the user, if the query is 
not clear, it can 
also correct the query by using the spell checker and reconfirm the query from the user to make the process precise enough to provide 
better results. The agent can also highlight 
specific information in
long documents to help the user to direct their 
attention to potential important content.

The user always has the
option to interrupt the ongoing communication process by entering a
new query directly into the Query Box. The communication finishes 
by the user 
ending the search
with success 
or with failure to address their information need. 

\subsubsection{System Workflow}
The system workflow is divided into two sections: Conversation Management and Search Management, dicussed in detail in our previous work. 

\begin{enumerate}
\item Conversation Management:
\label{Conversation Managment}This includes a Dialogues Manager, a Spell Checker and connection to the Wikipedia API. The Dialogue Manager validates the user input and 
either sends it to the AIML scripts or self-handles it, if 
the user input misspelt or incorrect. We use AIML scripts to implement the response to the user.
The system response to user input 
directed to the AIML scripts 
is determined by 
the AIML script, which can further classify the user's intent. The two major categories of intent are: {\em greeting\/} and {\em search\/}. The greeting intent is responsible for initializing, ending the conversation and system revealment. The search intent is responsible for directing the user input to the spell checker or wikipedia API and transferring
control to search management. The Spell Checking module is responsible for checking the spelling of the query and asking for suggestions from the user (for an example: If the user searches for ``viusal'' then the system would ask: Do you mean "visual"?). Once the user confirms ``yes'' or ``no'', then the query is forwarded to the Wikipedia API. 

\item Search Management:
\label{Search Management}
This is responsible for search and display of the top 3 search results. 
The user may also look for more sub-sections from a selected document. 
Search management also has an 
option to display 
the full document. This opens a display 
with important sections with respect to the query highlighted. The criteria for an important section is based on a Custom Algorithm which extracts important sentences based on a TF-IDF score for each sentence by selecting the top-scoring sentences. The top 30\% of extracted sentences are divided into clusters 
by Density based Clustering (DBSCAN)
to extract  diverse segments (combination of the sentences). Important segments are selected from these segments by using a cosine similarity score with the query. 

\end{enumerate}



\subsubsection{User Engagement}
The user can 
interact with both the search agent assistant and directly with the search engine. 
If the user commences a search from the Retrieval Results box,
the assistant initiates a dialogue to assist them in the search process. The system also provides support 
to the user in reading
full documents. As described above, important sections in long documents are highlighted to ease reading and reduce cognitive effort. 

\subsubsection{Review of Long Documents}

In our
study 
reported in \cite{Abhi}, we note users can spend 
considerable time reviewing 
long documents.
Our MCSI aims to
support these users and reduce their required effort by highlighting important segments with respect to the user's query, as described above.
This facility also provides the user with the opportunity to explore subsections within a document instead of needing to read a full long document. 

\subsubsection{Conventional Interface}
To enable direct comparison with our MCSI, a CSI for our study was formed by using the MCSI 
with the agent panel removed and the document highlighting facilities disabled. The searcher enters their query in the query box, 
document summaries are returned by the Wikipedia API, and full documents can be selected for viewing. 

\subsection{Information Needs for Study}


\begin{figure}[t] 
\vspace{-2mm}
\begin{flushleft}
\small
    {\tt It is late, but you can't get to sleep because a sore throat has taken hold and it is hard to swallow. You have run our of cough drops, and wonder if there are any folk remedies that might help you out until morning.}
\end{flushleft}
    \caption{Example backstory from UQV100 test collection.}
    \label{backstory}
   \vspace{-1mm}
\end{figure}


For our investigation, 
we wished to give searchers realistic information needs which could be satisfied using a standard web search engine. In order to control the form and detail of these, we decided to use a set of information needs specified within {\em backstories\/}, e.g. as shown in Figure \ref{backstory}. The backstories that we selected were taken from the UQV100 test collection \cite{Bailey:2016:UTC:2911451.2914671}, whose 
cognitive complexity is based on the Taxonomy of Learning \cite{krathwohl2002revision}. We decided to focus on the most cognitively engaging backstories, {\em Analyze\/} type, in the expectation that these would require the greatest level of user search engagement 
to satisfy the information need. 




Since the UQV100 topics were not provided with type 
labels, we selected a suitable subset as follows. The UQV100 topics were provided labeled with estimates of the number of queries which would need to be entered and the number of documents that would need to be accessed in order to satisfy the associated information need. We used the product of these figures as an estimate of the expected cognitive complexity, and then manually selected 12 of the highest scoring backstories that we rated as the most suitable for use by general web searchers, e.g. not requiring specific geographic knowledge or of specific events.

\subsection{Experimental Procedure}
Participants in our study had to complete search tasks based on the backstories using the MCSI and CSIs. Sessions were designed to assign search tasks and use of the alternative interfaces arranged to avoid potential sequence-related biasing effects. Each session consisted of multiple 
backstory search tasks.
While undertaking a search session, participants were required to 
complete a pre- and post task search questionnaires.
In this section we first give details of the practical experimental setup, then outline the questionnaires, and finally describe a pilot study undertaken to finalise the design of the study.


\subsubsection{Experimental Setup}
Participants used a setup of two computers arranged with two monitors side by side on a desk in our laboratory. One monitor was used for the search session, and the other to complete the online questionnaires. Participants carried out their search tasks accessing 
the Wikipedia search API using our interfaces running using a Google chrome browser. In addition, all search activities were recorded using a standard screen recorder tool to enable post-collection review of the user activities. Approval was obtained from our university Research Ethics Committee prior to undertaking the study. 
Participants were given printed 
instructions for their search sessions. 
each task. 

\subsubsection{Questionnaires}
Participants completed two questionnaires for each search task.
The questionnaire was divided into three sections:
\begin{itemize}
\item{{\em Basic Information Survey\/}: Participants entered their assigned user ID, age, occupation and task ID.}
\item{{\em Pre-Search\/}: Participants entered details of their pre-existing knowledge with respect to topic of the search task to be undertaken.}
\item{{\em Post-Search\/}: Post-search feedback from the user including their search experience, knowledge gain, and writing the post-search summary.}
\end{itemize}

The questionnaire was completed online in a Google form.

\subsubsection{Pilot Study}
A pilot study was conducted with two undergraduate students in Computer Science using two additional backstory search tasks. This enabled us to see how long it took them to complete the sections of the study using the CSI
and the MCSI, to gain insights into the likely behaviour of participants, and to generally debug the experimental setup.

Each of the pilot search tasks took around 30 minutes to complete. Feedback from the pilot study was used to refine the specification of the questionnaire. Results from the pilot study are not included in the analysis.

\begin{table}[t]
\centering
\tiny
\setlength{\tabcolsep}{3pt} 
\begin{tabular}{c|cccc}
\hline  \hline
\textbf{Task Load} &
  \textbf{CSI} &
  \textbf{MCSI} &
  \textbf{Percentage} &
  \textbf{P-Value} \\ 
  \textbf{Index} &
  
   \textbf{Mean} & 
   \textbf{Mean} &
  \textbf{Change} &
  \\ 
  \hline
  \hline
Mentally Demanding &
  4.16 &
  3.68 &
  11.54 &
   .273795 \\ \hline
Physically Demanding &
  3.12 &
  2.76 &
  11.54 &
.441676 \\     \hline
Hurried or Rushed &
  3.34 &
  2.76 &
  14.81 &
   .213878\\     \hline
Successful Accomplishing &
  4.28 &
  5.32 &
  -24.3 &
   .016199 \\     \hline
How hard did you have  & & & & \\ to work to accomplish? &
  4.44 &
  3.96 &
  10.81 &
  .270243 \\     \hline
How insecure, discouraged, irritated, & & & & \\ stressed, and annoyed were you?&
  3.32 &
  2.40 &
  27.71 &
.071443 \\ \hline  \hline
\end{tabular}
\caption{Task load index to compare the load on user while using both the systems (MCSI and CSI) with independent T two tailed test .}
\label{Task Load Index}
\end{table}

\subsubsection{Study Design}
Based on the result of the pilot study, each participant in the main study was assigned two of the selected 12 search task backstories with the expectation that their overall session would last around one hour. Pairs of backstories for each session were selected using a Latin square procedure. After every six tasks the sequence of allocation of the interface was rotated to avoid any type of sequence effect \cite{bradley1958complete}.

Each condition was repeated 
4 times with the expectation that this would give sufficient results to be able to observe significant differences where these are present. 
Since there were 12 tasks, this required 
24 subjects to participate in the study. 
In total, 27 subjects (9 Females, 18 Males) in the age group of 18-35 participated in our study  (excluding the pilot study), we examined
the data of 25 subject, since
2 subjects were found not to have followed
the instructions correctly. The study was
conducted in two phases. 
Each 
user had to perform a different
search task using the CSI
and MCSI with the sequencing of their use of the interfaces varied to avoid learning or biasing effects.

As well as completing the questionnaires, the subjects also attended a
semi-structured interview after completion of their session of two tasks
using both interface conditions.
The user actions in the videos and interviews were thematically labelled 
by two independent analysts
and Kappa coefficients were
calculated (approx mean .85) \cite{Kappa}. 
Disparities in labels were
`
resolved by mutual agreement between the 
analysts. 
The interview questionnaire dealt with user search experience, software usability and cognitive dimensions, and
was
quantitatively analyzed.
Based on the interview analysis, out of 25 participants 92\% 
were happy and satisfied with the MCSI. In all conditions, subjects preferred the MCSI. 
Showing that there is no sequence effect arising from the order of the interfaces in the search sessions. 


Each hypothesis of the study was 
tested using a T-Test (since
the number of samples was less than 31). Each hypothesis was
evaluated on a
number of factors which contribute to the examination in each dimension as discussed below. 

              



\section{Study Results}
\label{procedure}
The MCSI was compared with the conventional interface 
using an implicit evaluation method examining multiple dimensions: cognitive load, knowledge gain, usability and search satisfactions.  


\begin{table}[t] 
\centering
\tiny
\setlength{\tabcolsep}{3pt} 
\begin{tabular}{c|ccccc}
\hline  \hline
\textbf{Topic} &
  \textbf{CSI} &
  \textbf{MCSI} &
  \textbf{Percentage} & \textbf{P value} \\ & \textbf{Mean} & \textbf{Mean} & \textbf{Change} &
   \\ \hline   \hline
Easy to use\textsuperscript{*}&
  4.04 &
  5.96 &
  47.52 &
.000059 \\ \hline
Simple to use  &
  4.48 &
  5.92 &
  32.14 &
.003526 \\  \hline
Effectively complete my work\textsuperscript{*} &
  3.92 &
  5.64 &
  43.88 &
  .000226\\  \hline
Quickly complete my work\textsuperscript{*}  &
  3.72 &
  5.76 &
  54.84 &
.00003 \\  \hline
Efficiently complete my work\textsuperscript{*}  &
  3.88 &
  5.76 &
  48.45 &
.000045 \\  \hline
Comfortable using this system\textsuperscript{*} &
  4.16 &
  5.88 &
  41.35 &
.000471 \\  \hline
Whenever I make a mistake using the & & & & \\ system, I recover easily and quickly\textsuperscript{*} &
  4.04 &
  5.44 &
  34.65 &
.006827\\  \hline
The information is clear\textsuperscript{*} &
  4.16 &
  5.92 &
  42.31 &
  .000072 \\  \hline
It is easy to find the information I needed\textsuperscript{*} &
  4.00 &
  5.48 &
  37 &
.000706\\  \hline
The information is effective in & & & & \\ helping me complete the tasks and scenarios\textsuperscript{*} &
  4.20 &
  5.68 &
  35.24 &
  .000675. \\  \hline
The organization of information & & & & \\ on the system screens is clear\textsuperscript{*} &
  4.44 &
  5.92 &
  33.33 &
.000184\\  \hline
The interface of this system is pleasant\textsuperscript{*} &
  4.28 &
  6.08 &
  42.06 &
  .00002\\  \hline
Like using the interface\textsuperscript{*} &
  4.20 &
  6.12 &
  45.71 &
 .000014\\  \hline
This system has all the functions  &  & & & \\
 and capabilities I expect it to have\textsuperscript{*} & 4.08 &
  5.72 &
  40.2 &
   .000168 \\  \hline
Overall, I am satisfied with this system\textsuperscript{*} &
  4.16 &
  5.92 &
  42.31 &
 .000029\\ \hline  \hline
\end{tabular}
\caption{Post Study System Usability Questionnaire (PSSUQ).}
\label{Post-Study System Usability Questionnaire}
\end{table}

\subsection{Cognitive dimensions}

Conventional search can impose a significant cognitive load on the searcher \cite{kaushik2019dialogue}. 
An important factor in the evaluation of conversational systems is measurement of the cognitive load experienced by users.
To measure user 
workload, the NASA Ames Research Centre proposed the NASA Task Load Index \cite{hart1988development,Abhievaluation}.
In terms of cognitive load, the user was
asked to evaluate the conventional 
interface and MCSI 
in
6 dimensions from the NASA Task Load Index associated with 
mental load and physical load,
as shown in 
Table \ref{Task Load Index}.

\begin{enumerate}
    
    \item \textbf{HO: Users experience a similar task load during the search with multiple interfaces}:
    The grading scale of the NASA Task Load Index measure lie between 0 (low) - 7 (High). We 
    compared the mean difference of both 
    systems on all six parameters. In all 
    aspects, subjects experienced lower
    task load using the MCSI.
    Subjects claimed more success in accomplishing the task using the MCSI.
    Results for accomplishing the task were 
    statistically significantly different. 
    Subjects felt less insecure, discouraged, irritated, stressed, and annoyed, while using the MCSI
    with a
    significant difference (P$<$0.10). This implies that the null hypothesis was rejected on the basis of the Task Load index. Although 
    four factors were not significantly different, 
    the mean difference between both the systems on these factors was more than 
    10\%. 
    This shows 
    that the user experienced less subjective mental workload while using the MCSI. 
\end{enumerate}


\begin{table}[t] 
\small
    \centering
    \begin{tabular}{l|l}
    \hline
    Parameter & Definition \\
    \hline
   Dqual & Comparison of the quality of facts \\ & in the summary in range 0-3 where 0 \\ 
   & represents irrelevant facts and  \\ &  3 specific details with relevant facts. \\
   \hline 
   Dintrp & Measures the association of facts in  \\ &  a summary in the range 0-2 where \\ 
   & 0 represents no association of the facts and \\ &  2 that all facts \\ & in a summary are associated  \\ &  with each other in a meaning. \\
      \hline 
   Dcrit & Examines the quality of critiques of topic   \\ & written by the author \\ &  in range the 0-1 where 0 \\
   & represents facts are listed with  \\ &  without thought or analysis  \\ & of their value and 1 \\
   & where both advantages and  \\ &  disadvantages of the facts are given. \\
   \hline
    \end{tabular}
    \caption{Summary Comparison Metric \cite{wilson2013comparison}}
    \label{Summary Comparison Metric}
\end{table}

\subsection{Usability}
CS
studies generally do not explore the dimensions of 
software usability. However, it is important to understand the challenges and opportunities of CSs
on the basis of software requirements analysis. This allows a system to be evaluated based on real-life deployment and to identify areas  for improvement. Lower effectiveness and efficiency of a software system can increase cognitive load, reduce engagement and act as a barrier in the process of learning while searching \cite{Abhi,vakkari2016searching,kaushik2019dialogue}.
Usability is an
important evaluation metric of 
interactive software. IBM Computer Usability Satisfaction Questionnaires are
a Psychometric Evaluation for 
software from the perspective of the user \cite{lewis1995ibm} known as the Post-Study System Usability Questionnaire (PSSUQ) Administration and Scoring. The PSSUQ was
evaluated using
four dimensions:
overall satisfaction score (OVERALL), system usefulness (SYSUSE), information quality (INFOQUAL) and interface quality (INTERQUAL), which include fifteen parameters. On each dimension, the MCSI 
outperformed the CSI. The grading scale 
lies between 0 (low) - 7 (High).  We 
compared the mean difference of both 
systems on all parameters. In all 
aspects, subjects experienced less task load when using the MCSI ,
as shown in Table \ref{Post-Study System Usability Questionnaire}. 

\begin{enumerate}
    \item \textbf{
H0: User Psychometric Evaluation for the conversational
interface and 
conventional
search has no significant difference} 
A T Independent test was
conducted. 
It was
found that for
all the parameters the 
MCSI outperformed the CSI.
The null hypothesis was
rejected and the H1 hypothesis was accepted, which is that the MCSI 
performs better than the CSI.
\end{enumerate}

\begin{table}[t] 
\centering
\tiny
\begin{tabular}{cccc}
\hline
\textbf{Topic}                 & \textbf{Pre-Task} & \textbf{Post Task} & \textbf{P Value} \\ \hline
DQual (1-3)\textsuperscript{*}            & 0.32     & 1.56      & .00005  \\
DCrit (0-1)       & 0        & 0.32      & .0026   \\
DIntrp (0-2)\textsuperscript{*} & 0        & 0.84      & .00005 \\ \hline
\end{tabular}

\caption{Comparison of Pre-search and Post-search summary for the CSI (Change in Knowledge).} 
\label{ComparisonT}

\vspace{2ex}

\tiny
\begin{tabular}{llll}
\hline
\textbf{Topic}                  & \textbf{Pre-search} & \textbf{Post search} & \textbf{P Value}          \\ \hline
DQual (1-3)\textsuperscript{*}            & 0.52       & 2.12        & \textless .00001 \\
DCrit (0-1)\textsuperscript{*}        & 0.12       & 0.72        & \textless .00001 \\
DIntrp (0-2)\textsuperscript{*} & 0.28       & 1.36        & \textless.00001   \\ \hline       
\end{tabular}

\caption{Comparison of Pre-search and Post-search summary for the MCSI (Change in Knowledge).} 
\label{ComparisonI}


\vspace{2ex}

\tiny
\begin{tabular}{lll} \hline
\textbf{Parameters}                 & \textbf{Influence} & \textbf{P value ($<$ 0.5)}                                                \\ \hline
Increase in Critique       & 87\%                & .048153 \\
Increase in Quality        & 29\%                & .299076                                 \\
Increase in Interpretation & 22\%                & .312712.        \\
\hline
\end{tabular}
\caption{Comparison of Traditional Search and Interface Search}
\label{ComparisonB}
\end{table}

\subsection{Knowledge Expansion} 
Satisfaction of the user's information need is directly related to their knowledge gain about the search topic. Knowledge gain can be measured based on recall of new facts gained after the completion of the search process \cite{wilson2013comparison}. 
We investigated knowledge expansion 
using
a comparison of 
pre-search 
and post-search summaries written by the participant, based on a number of parameters, as shown in Table \ref{Summary Comparison Metric}, while using both the systems. We divide the hypothesis into two sub-parts as follows: 

\begin{enumerate}
    \item Comparison of
    pre-search and post-search summaries: This is to verify the knowledge expansion after each task independent of the search interface used by the participant.
    \item Comparison of the mean difference between
    pre-search and post-search summaries for 
    each interface: This is to verify which interface supported users better in gaining 
    knowledge. 
\end{enumerate}

The user gains knowledge during the search when using either of 
the search interfaces.  
To measure their knowledge gain, we 
asked subjects to write a short summary of the topic before the search and after the search. Each 
summary was
analyzed based on three 
criteria as described in
\cite{wilson2013comparison}: Quality of Facts (DQual), Intrepterations (DInterpretation) and Critiques (DCritique), as shown in Table \ref{Summary Comparison Metric}. 
The summaries were
scored against
these three factors by two independent analysts 
with the Kappa coefficient (Approx .85) \cite{Kappa}. We conducted 
hypothesis T dependent testing on tasks completed using both the 
conventional
search interface
and the MCSI.

\begin{enumerate}
    \item \textbf{H0: No significant difference in the increase of the knowledge after completing the search task in both 
    settings}: As shown in Tables \ref{ComparisonT} and \ref{ComparisonI}, the pre-search score and post-search score for
    all three factors were statistically significant in both the search settings. This implies that subjects expand their knowledge while carrying out the search.  This rejects the null hypothesis which leads to the alternative hypothesis which concludes that users experienced significant increase in their knowledge after search in both search settings.
\end{enumerate}

After concluding the alternative hypothesis, it was 
important to investigate whether one
system was better in 
expanding the user's 
knowledge. We purposed and tested the following hypothesis.

\begin{enumerate}
    \item \textbf{H0: Knowledge gain during the search is independent of the interface design:} In this test, we compared the Mean of the difference in the 
    score for
    pre-search and post-search summaries in both 
    settings. 
    conducted on the change of the three parameter scores as discussed above for the hypothesis testing as shown in the Table \ref{ComparisonB}. It was
    found that in the MCSI interface setting, the subjects scored higher in the change of critique, quality and interpretation. This implies that the subjects learned more while using the 
    MCSI. The difference in critique score was statistically significant, while the other two parameters were not statistically significant.
    The quality and interpretation 
    increased more than 20\% while using the MCSI.
    This confirms
    the alternative hypothesis, subjects' knowledge expands more 
    when using the MCSI.
\end{enumerate}

\begin{table}[!ht]
    \centering
    \tiny
    \setlength{\tabcolsep}{3pt} 
    \begin{tabular}{l|l|l|l|l}
\hline \hline 
\textbf{Parameters} & \textbf{CSI} & \textbf{MCSI} & \textbf{Percentage} & \textbf{P value}                                      \\ 
& \textbf{Mean} & \textbf{Mean} & \textbf{Change} & \\ 

\hline \hline 

Difficulty in finding the & & & & \\ information needed to  & & & & \\ address this task?           & 4.64    & 3.16            & -35.25     & .002168         \\ \hline 
Quality of text presented & & & & \\  with respect to your information  & & & & \\ need and query?                                                                  & 4.52    & 5.64           & 21.55      & .010465      \\ \hline 
How useful were the search & & & & \\  results in the whole search task?                                                                                                                                    & 4.04   & 5.12            & 23.08      & .029826      \\ \hline 
How useful was the text shown & & & & \\  in the whole search task in  & & & & \\ satisfying the information need?                                                                                                      & 4.08     & 5.36           & 27.62      &  .010245      \\ \hline 
Did you find yourself to be cognitively & & & & \\  engaged while carrying  & & & & \\ out the search task? \textsuperscript{*}                                                                                                             & 3.92      & 5.92             & 42.31      & .000015      \\ \hline 
Did you expand your knowledge about & & & & \\  the topic while completing  & & & & \\ this search task?                                                                                                                & 4.84     & 6           & 20         & .005026     \\ \hline 
I feel that I now have a better & & & & \\  understanding of the topic  & & & & \\ of this search task.                                                                                                                 & 4.56     & 5.88            & 25.64      &  .002094      \\ \hline 
How would you grade the success & & & & \\  of your search session  & & & & \\ for this topic?                                                                                                                          & 4.48     & 5.72           & 24.35      & .005937      \\ \hline 
How do you rate your assigned & & & & \\  search setting in terms of  & & & & \\ understanding your inputs?                                                                                                            & 3.72      & 5.40            & 39.18      & .003121.          \\ \hline 
How do you rate your assigned & & & & \\  search setting in the presentation  & & & & \\ of the search results?\textsuperscript{*}                                                                                                         & 3.84     & 5.76            & 45.45      &.00001 \\ \hline 

How do you rate the suggestion(s) & & & & \\  skills of your assigned  & & & & \\ search setting?\textsuperscript{*}                                                                                                                       & 3.72      & 5.56            & 54.44      &  .000053     \\   \hline               
\end{tabular}
\caption{Characteristics of the search process \cite{vakkari2016searching} by the change in knowledge structure where * indicates statistically significant results.}
\label{Flowchart of characteristic of search process by the change in knowledge structure}
\vspace{-2ex}
\end{table}

\subsection{Search Experience}
Learning while searching is an integral part of the information seeking process. Based on the search as learning  proposed by Vakkeri \cite{vakkari2016searching}, the user search experience 
can be 
evaluated on 15 parameters, including
the relevance of the search result, the quality of the text presented by the interface, and understanding of the topic in both the search settings via pre-search and post-search questionnaires.
\begin{enumerate}
    \item \textbf{H0: Subjects find no significant difference between while using both the interfaces:} The T-independent test was
    conducted among
    all 15
    parameters, shown in Table \ref{Flowchart of characteristic of search process by the change in knowledge structure}. 
    It was 
    found that the null hypothesis was 
    rejected 
   Subjects search experience was statistically significantly better with the MCSI. 
In the pre-search questionnaire, subjects were asked to anticipate the difficultly level of the search before starting the search and in post-search questionnaire, subjects were asked to indicate
the difficulty level they actually experienced. 
It was
observed that 
pre-search anticipated difficulty level and the post-search actual difficulty level 
increased for 
the CSI 
(16\%) and decreased in the case of MCSI 
search task (14\%).

\end{enumerate}





\subsection{Interactive User Experience}


\begin{table}[t] 
\centering
\tiny
\setlength{\tabcolsep}{4pt} 
\begin{tabular}{llllll}
\hline 
\textbf{Negative}   & \textbf{Positive}     & \textbf{Scale}             & \textbf{CSI\_Mean} & \textbf{MCSI\_Mean} & \textbf{P\_Values} \\ \hline 
obstructive     & supportive   & P & 3.44     & 5.60   & 2.96e-08 \\ 
complicated     & easy         & P & 3.40    & 5.76    & 7.84e-09 \\ 
inefficient     & efficient    & P & 2.88   & 4.40   & 1.69e-05  \\ 
confusing       & clear        & P & 3.40   & 5.48    & 2.31e-06 \\ 
boring          & exciting     & H   & 2.64   & 5.44    & 8.88e-16 \\ 
not interesting & interesting  & H   & 2.48    & 5.48     & 9.76e-15 \\ 
conventional    & inventive    & H   & 2.36     & 6.28    & 1.17e-14 \\ 
usual           & leading edge & H & 1.96     & 5.20     & 8.95e-12 \\ 
\hline
\end{tabular}
\caption{UEQ-S score based on CSI and MCSI where 'P' stands for Pragmatic Quality and 'H' stands for Hedonic Quality (statistically significant).}
\label{CHap7:USQ_S}
\end{table}

To ensure a conversational search system provides reasonable User Experience (UX), it is critical to have a measurability which defines user insights about the system. A UX questionnaire for interactive products is the User Experience Questionnaire (UEQ-S) \cite{laugwitz2008construction,schrepp2017design,hinderks2018benchmark}. 
This questionnaire also enables 
analysis and interpret outcomes by comparing with benchmarks of a larger dataset of outcomes for other interactive products \cite{hinderks2018benchmark}. This questionnaire also provides the opportunity to compare interactive products with each other. 
For specified purposes, a brief version (UEQ-S) was prepared which had only 8 parameters to be considered \cite {hinderks2018benchmark}. UEQ-S was preferred for the MCSI, since
it is mostly used for interactive products. For example, users filled the experience questionnaire after finishing the search task, if there were too many questions, a user may not complete the answers fully or even refuse to complete 
it (as they have finished the search task and are in the process of leaving or starting the next task, so the motivation to invest more time on feedback may be limited).
The UEQ-S contains two meta dimensions Pragmatic and Hedonic quality. Each dimension contains 4 different parameters, as shown in 
Table \ref{CHap7:USQ_S}. Pragmatic quality explores the usage experience of the search system, while Hedonic quality explores the pleasantness of use of the system.

\begin{enumerate}
    
    \item \textbf{HO: Users feel a similar interactive experience when using the different interfaces}:
    Users evaluated the system based on 8 parameters as shown in Table \ref{CHap7:USQ_S}. The grading scale was assigned between 0 (low) - 7 (High). We 
    compared the mean difference of both 
    systems on all parameters. In all 
    aspects, subjects experience was positive in Pragmatic quality and Hedonic quality when using the MCSI, 
    and statistically significantly different in comparison to the CSI. 
    Subjects felt obstructive, complicated, confusing, inefficient, and boring, while using the CSI with significant difference (P$<$0.10). This implies that the null hypothesis was rejected on the basis of the user experience. 
    Based on these findings, we can conclude that the user experience was more pleasant and easy while using the MCSI. 
\end{enumerate}

\subsection{Analysis of Study Results}

In summary, 
hypothesis testing showed that 
the MCSI 
reduced cognitive load, increased knowledge expansion, increased cognitive engagement and 
provided a better search experience load. 
Based on the results of the study, a number of research questions dealing with factors relating to 
conversational search, 
the challenges of 
conventional
search, and 
user search behaviour can be addressed. 

\subsubsection{RQ1: What are the factors that support search using the MCSI.}

Around 92\% of the subjects claim in the post-search interview that the MCSI
was better than
the CSI.
Around 48\%
found
that the 
MCSI
allowed them to more easily access the information. A similar view was
found in terms of information relevance and its structure as presented to the user.  Around 38\% of subjects were satisfied with the options and suggestions provided by the MCSI.
The other reasons for their satisfaction were the highlighting of segments in long documents, finding the 
search system effective, its being interactive and
engaging, and user friendly.

\subsubsection{RQ2: What are the challenges with the conventional
search system?}
\vspace{-1ex}
Subjects found
some major challenges in 
completing the search tasks
with the CSI.
The limitations were mainly
based on observations from 
user interactions and 
feedback after the search task. The limitations can be
divided into five broad categories.

\noindent \textbf{Exploration}: 
Around 60\% of the subjects claimed they found
it difficult to explore the content with the CSI, which meant that they were unable
to learn through the search process. It was noted
that they needed to expend much
effort to go through whole documents, which discouraged them from 
exploring further to satisfy their information need. Another reason was that too much information was displayed to them 
on the page which confused them
during 
the process of information seeking.

\noindent \textbf{Cognitive Load}: Around 28\% of subjects experienced issues with
cognitive load using
the CSI.
In current search systems, a query to 
the search engine returns the best document in a
single shot. The user may need to perform multiple searches by modifying the search query each time to satisfy their information need. There are multiple
limitations 
associated with this
single query search approach which put 
high cognitive load on the user. The following points highlight the limitations and weaknesses of single-shot search \cite{kaushik2019dialogue}. 

\begin{enumerate}
    \item The user must completely describe their information need in a single query. 
    \item The user may not be able to adequately describe their information need.
    \item High cognitive load on the user in forming a query.
    \item An information retrieval system should return relevant content in a single pass based on the query.
    \item The user must inspect returned content to identify
    relevant information.
\end{enumerate}

\noindent \textbf{Interaction and Engagement}: 
8\% found 
difficulty in engaging and interactive with long documents.
Subjects can find
content in long documents irrelevant or vague with respect to their specific information need.  Using 
the CSI,
32\% of the subjects did not
find the long documents precise enough to satisfy their information need. In contrast, 90\% of them
were satisfied with the way information was presented to them in the MCSI,
although the Wikipedia
API and underlying retrieval method was same for both interfaces. 

\noindent \textbf{Highlighting}: Another issue 
which was
referred to by around 8\% of subjects related to text highlighting.
Subjects found
that the absence of 
highlighting
in the CSI
was frustrating.


\subsubsection{RQ3: Does Highlighting important segments support users in effective and efficient search?, and Why?}
92\% of subjects liked the document highlighting
options in the MCSI.
The following 
reasons were identified for 
choosing this. 

\begin{enumerate}
    \item Interactive and Engaging: Around 28\% of subjects claimed that they were able to engage and interact with documents 
    better 
    by using the highlighting options.
    \item Helpful: 68\% of the subjects found
    highlighted documents helpful in information seeking.  
    \item Reduce the Cognitive Load:  Around 24\% of the subjects believed that the highlighted documents reduced their cognitive load.
    \item Access to Relevant information: 
    36\% of the subjects believed that 
    highlighted documents helped them to more easily access useful
    information.
\end{enumerate}

\begin{table*}[!ht]
\centering
\tiny
\begin{tabular}{lccccccc} \hline
\multicolumn{7}{l}{Confidence   intervals (p=0.05) per scale} \\ \hline
Scale & Mean (-3 to 3) & Std. Dev. & N & C & \multicolumn{2}{c}{C interval} & alpha value\\ \hline
P & -0.720 & 1.349 & 25 & 0.529 & -1.249 & -0.191 & 0.91 \\
H & -1.640 & 1.233 & 25 & 0.483 & -2.213 & -1.157 & 0.92  \\
Overall & -1.180 & 1.207 & 25 & 0.473 & -1.653 & -0.707 & 0.91 \\ \hline
\end{tabular}
\caption{CSI confidence intervals on UEQ-S where, 'P' stands for Pragmatic Quality, 'H' stands for Hedonic Quality and 'C' stands for Confidence.}
\label{7ConfidenceCSI}
\end{table*}

\begin{table*}[!ht]
\centering
\tiny
\begin{tabular}{lccccccc}
\hline
\multicolumn{7}{l}{CSI   Confidence intervals (p=0.05) per scale} \\ \hline
Scale (-3 to 3) & Mean (-3 to 3) & Std. Dev. & N & C & \multicolumn{2}{c}{C Interval} & alpha value \\ \hline
P & 1.310 & 0.596 & 25 & 0.234 & 1.076 & 1.544
 & 0.79\\
H & 1.600 & 0.559 & 25 & 0.219 & 1.381 & 1.819 & 0.79 \\
Overall & 1.455 & 0.519 & 25 & 0.203 & 1.252 & 1.658 & 0.79 \\ \hline
\end{tabular}
\caption{MCSI confidence intervals on UEQ-S, where 'P' stands for Pragmatic Quality,'H' stands for Hedonic Quality and 'C' stands for Confidence.}
\label{7ConfidenceMCSI}
\end{table*}

\subsubsection{RQ4: What are the challenges and opportunities to support exploratory search in 
conversational settings?} 

The 
majority of subjects (92\%) 
claimed that the MCSI
was better. The remaining subjects (8\%) faced some
challenges using it.
Subjects wanted 
more sections and subsections in
the documents to support their exploration, and 
also wanted 
support of
image search. Around 4\% of the subjects felt the need for 
improvement in
operational speed and 
better incorporation of standard features such as spellchecking. Subjects found
the chat interface helpful for 
exploring 
long documents. They were keen to see the addition of 
speech as a mode of user interaction and a more refined 
algorithm for the selection of 
images for 
presentation to the user. 

Subjects appreciated the 
usefulness of the interface 
in 
supporting
exploratory search, but suggested that this would be further improved by the incorporation of 
a question answering facility.

\subsubsection{RQ5: How does user experience vary between search settings in comparison to each other?}


\begin{enumerate}
    \item \textbf{Observing the Pragmatic and Hedonic properties of CSI}: The users provided feedback based on their experience using the CSI. 
    The CSI score is negative with respect to both Pragmatic and Hedonic properties and the overall score is also negative.  From this 
    we can infer that the user’s experience of the CSI system is neither effective nor efficient, as shown in the Table \ref{7ConfidenceCSI}. From Table \ref{7ConfidenceCSI}, we can calculate the mean range after data transformation for UEQ-S where is -3 too negative and +3 is too positive. Table \ref{7ConfidenceCSI} shows the confidence interval and confidence level. The smaller the confidence interval the higher the precision \cite{UEQ}. The confidence interval and confidence level confirm our analysis that all the dimensions of Pragmatic and Hedonic properties were negatively experienced by the users. Generally, items belonging to the same scale should be highly correlated. To verify the user consistency, alpha-coefficient correlation was calculated using the UEQ-S toolkit. As per different studies, an alpha value $>$ 0.7 is considered sufficiently consistent \cite{hinderks2018benchmark}. This shows that user marking of the CSI
    is consistent. The UEQ-S tool kit also provides an option to detect random and non-serious answers by the users \cite{UEQ} \cite{hinderks2018benchmark}. This is carried out by checking how much the best and worst evaluation of an item in a scale differ. Based on this evaluation, the users' feedback does not show any suspicious data.
   
\item \textbf{Observing the Pragmatic and Hedonic properties of MCSI}: 
The MCSI scored positive in Pragmatic, Hedonic and Overall score from which we can infer that the user’s experience of the MCSI is good in general and with good ease of use. Table \ref{7ConfidenceMCSI} shows the confidence interval and confidence level. The confidence interval and confidence level confirms our analysis that all the dimensions of pragmatic and hedonic scores were positively experienced by the users. Alpha-coefficient correlation \cite{hinderks2018benchmark} confirms that the marking of MCSI by the users is consistent. The UEQ-S toolkit also provides an option to detect random and non-serious answers by users. This is conducted by checking how much the best and worst evaluation of an item in a scale differ. Based on this evaluation, the users' feedback does not detect any suspicious data.
\end{enumerate}


\subsubsection{RQ6: How does user experience vary for both search settings in comparison to a standard benchmark?}

\begin{enumerate}
\item \textbf{Comparison of the CSI with the standard benchmark}: This benchmark was developed based users on feedback on 21 interactive products \cite{hinderks2018benchmark}. Based on the comparison from the benchmark, the CSI UX is far below the mean of the interactive products (Pragmatic Quality $<$ 0.4, Hedonic Quality $<$ 0.37 and overall $<$ 0.38). 
This signifies that the UX with the CSI needs major improvement on Pragmatic and Hedonic sectors. In the comparison to the benchmark, the CSI rates as a low quality of user experience and lies in the range of worst 25\% of the products.


\item \textbf{Comparison of the MCSI with the standard benchmark}: Based on the comparison from the benchmark \cite{hinderks2018benchmark}, the MCSI UX is far above the mean of the interactive products (Pragmatic Quality $>$ 0.4, Hedonic Quality $>$ 0.37 and overall $>$ 0.38). 
This signifies the UX of the MCSI compared to other interactive products (benchmark) is very high and is of excellent level, and lies in the range of 10\% best results. 

\end{enumerate}

\section{Conclusions and Observations}
\label{conclude}


The study reported in this paper 
indicates that subjects found our MCSI more helpful than a closely matched CSI.
We also observed types of user behaviour while using MCSI which are
different to those when using
a CSI.
Using our agent-based system, we observe the natural expectations of user search in conversational settings. We observed 
that subjects do not encounter any difficulty in using the new interface, because it seems to be similar to the standard search interface with the additional capabilities of conversation.  We also observe that the information space and its structure is a key component in information seeking. Subjects found highlighting important segments in long documents enables them to access information much easily. The MCSI made the search process less cognitively demanding and more cognitively engaging. 



Clearly our
existing rule-based search agent can be extended in terms of functionality, and going forward we aim to examine basing its functionality on machine learning based methods, but this will require access to sufficient suitable training data, which is not available at this prototype stage.

\section*{Acknowledgement}

This work was supported by Science Foundation Ireland as part of the ADAPT Centre (Grant 13\//RC\//2106) at Dublin City University.






\bibliographystyle{apalike}
\bibliography{references}  






\end{document}